\documentclass[aps,prl,twocolumn,floatfix,superscriptaddress]{revtex4-1}

\usepackage{graphicx}
\usepackage[colorlinks=true,linkcolor = blue,
            urlcolor  = blue,
            citecolor = blue,
            anchorcolor = black]{hyperref}
\usepackage{siunitx}

\begin{document}
\title{One-dimensional edge transport in few-layer $\mathrm{WTe_2}$}

\author{Artem Kononov}
\email{Artem.Kononov@unibas.ch}
\affiliation{\footnotesize Department of Physics, University of Basel, Klingelbergstrasse 82, CH-4056 Basel, Switzerland}
\affiliation{\footnotesize Institute of Solid State Physics of the Russian Academy of Sciences - Chernogolovka, Moscow District, 2 Academician Ossipyan str., 142432 Russia}

\author{Gulibusitan Abulizi}
\affiliation{\footnotesize Department of Physics, University of Basel, Klingelbergstrasse 82, CH-4056 Basel, Switzerland}

\author{Kejian Qu}
\affiliation{\footnotesize Department of Materials Science and Engineering, University of Tennessee, Knoxville, TN 37996, United States}

\author{Jiaqiang Yan}
\affiliation{\footnotesize Materials Science and Technology Division, Oak Ridge National Laboratory, Oak Ridge, TN 37831, United States}
\affiliation{\footnotesize Department of Materials Science and Engineering, University of Tennessee, Knoxville, TN 37996, United States}

\author{David Mandrus}
\affiliation{\footnotesize Department of Materials Science and Engineering, University of Tennessee, Knoxville, TN 37996, United States}
\affiliation{\footnotesize Materials Science and Technology Division, Oak Ridge National Laboratory, Oak Ridge, TN 37831, United States}

\author{Kenji Watanabe}
\affiliation{\footnotesize National Institute for Material Science, 1-1 Namiki, Tsukuba 305-0044, Japan}

\author{Takashi Taniguchi}
\affiliation{\footnotesize National Institute for Material Science, 1-1 Namiki, Tsukuba 305-0044, Japan}

\author{Christian Schönenberger}
\email{Christian.Schoenenberger@unibas.ch}
\affiliation{\footnotesize Department of Physics, University of Basel, Klingelbergstrasse 82, CH-4056 Basel, Switzerland}
\affiliation{\footnotesize Swiss Nanoscience Institute, University of Basel,
Klingelbergstrasse 82, CH-4056 Basel, Switzerland}

\begin{abstract}

$\mathrm{WTe_2}$ is a layered transitional metal dichalcogenide (TMD) with a number of intriguing topological properties. Recently, $\mathrm{WTe_2}$ has been predicted to be a higher-order topological insulator (HOTI) with topologically protected hinge states along the edges. The gapless nature of $\mathrm{WTe_2}$ complicates the observation of one-dimensional (1D) topological states in transport due to their small contribution relative to the bulk. Here, we study the behavior of the Josephson effect in magnetic field to distinguish edge from bulk transport.  The Josephson effect in few-layer $\mathrm{WTe_2}$ reveals 1D states residing on the edges and steps. Moreover, our data demonstrates a combination of Josephson transport properties observed solely in another HOTI – bismuth, including Josephson transport over micrometers distances, extreme robustness in magnetic field and non-sinusoidal current-phase relation (CPR). Our observations strongly suggest the topological origin of the 1D states and that few-layer $\mathrm{WTe_2}$ is a HOTI.

\end{abstract}
\maketitle

Materials with non-trivial topology attract a lot of attention due to their intriguing properties and potential to harness them for quantum computing. Non-abelian excitations, occurring when topology meets superconductivity, are especially interesting for applications~\cite{quantcomp}. Many realizations of these excitations have been proposed and implemented recently, including designing topological superconductivity by combining spin-orbit interaction and Zeeman effect with normal s-wave superconductors~\cite{Kitaev}, or by proximity inducing superconductivity in topological insulators~\cite{Fu}. Recently, it has also been demonstrated that one can engineer them in hinge states of a higher-order topological insulator (HOTI) combined with proximity induced superconductivity~\cite{Jaeck}. The layered TMD $\mathrm{WTe_2}$, which in the form of a 3D crystal is a Weyl semimetal~\cite{Soluyanov, Weyl_II} and a 2D topological insulator in the form of a monolayer~\cite{Fei, Wu}, has been predicted to be a HOTI~\cite{HOTI}, hosting topological hinge states on the edges and steps of the crystal. However, the bulk conductivity of $\mathrm{WTe_2}$ complicates the observation of these states. One way to overcome bulk conductivity is to use local measurement techniques such as scanning tunneling spectroscopy~\cite{Jaeck}. Another possibility is to employ the Josephson effect~\cite{Murani,Schindler,Shevtsov2}. Here, the evolution of the critical current $I_c(B_{\perp})$ with perpendicular magnetic field $B_{\perp}$ is connected with the current distribution in the plane by a Fourier transform~\cite{Dynes}. The asymmetry of the critical current can provide additional information about properties of the supercurrent carrying states. The asymmetric Josephson effect (AJE) is expected  in systems with a non-sinusoidal CPR~\cite{AJE}, which is often linked with the presence of Andreev bound states with high transmission~\cite{Sochnikov}. The AJE has been previously observed in a 2D topological insulator coupled to a superconductor~\cite{Bocquillon}.

Here, we reveal 1D states along edges and steps in few-layer $\mathrm{WTe_2}$ by studying the Josephson effect in a perpendicular magnetic field. The superconducting contacts required for Josephson junctions are realized by a lithographically patterned Pd film that is in contact with clean $\mathrm{WTe_2}$ and induces superconductivity therein. We found that a Josephson current can be measured over distances up to $3~\si{\micro\meter}$ and it withstands magnetic fields up to $2~\si{\tesla}$, suggesting its 1D nature with a very tight lateral confinement. Moreover, transport through these 1D states shows signatures of the asymmetric Josephson effect. We think that the observed behavior can be a result of Josephson transport through hinge states due to higher-order topology in $\mathrm{WTe_2}$.

\begin{figure}[htb]
    \centering
    \includegraphics[clip, trim=6.9cm 7cm 7cm 7cm,width=0.4\textwidth]{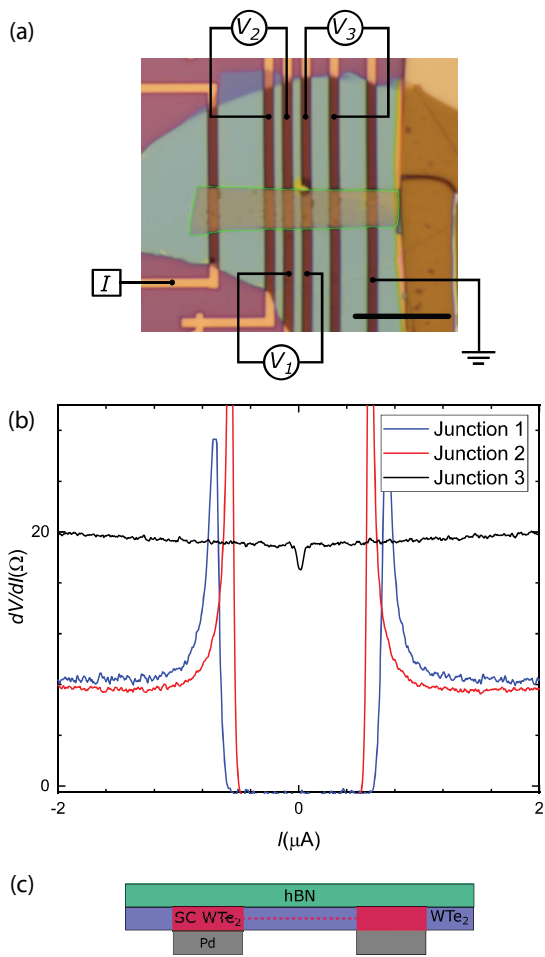}
    \caption{(\textbf{a}) Optical image of the device~1 (scale bar $10~\mu m$) with a sketch of the four-terminal measurement setup. (\textbf{b}) Four-terminal differential resistance $dV/dI$ of three junctions 1-3 with lengths $1, 1$ and $2~\si{\micro\meter}$, respectively. The $1~\si{\micro\meter}$ long junctions demonstrate zero differential resistance at small currents as a result of the Josephson effect. (\textbf{c}) Schematic side view of the sample illustrating its state: above the Pd leads $\mathrm{WTe_2}$ is superconducting (red regions) and the current between these regions is mediated by the Josephson effect (dashed lines).}
    \label{WTe02}
\end{figure}

Fig.~\ref{WTe02}(a) demonstrates an optical image of our first device. It consists of a few-layer ($\sim12$) thick $\mathrm{WTe_2}$ flake covered with hBN and placed on the prepatterned Pd leads on $\mathrm{SiO_2}$ substrate. The leads are forming several junctions with different lengths $1-4~\si{\micro\meter}$. We measured the differential resistances of the junctions in the four-probe setup sketched in Fig.~\ref{WTe02}(a). Additional details of the fabrication process and the measurement setup are provided in the supplementary information. All measurements were performed in the dilution refrigerator with base temperature of $30~\si{\milli\kelvin}$.

Fig.~\ref{WTe02}(b) demonstrates experimental $dV/dI(I)$ dependencies of three junctions of the device~1. The differential resistance goes to zero at small currents for the two $1~\si{\micro\meter}$-long junctions. For the $2~\si{\micro\meter}$ junction the differential resistance doesn't go to zero, but has a small dip at zero current. Similar results are obtained in all studied samples. Moreover, the observed behavior is present only below a certain temperature and magnetic field. This behavior is typical for Josephson junctions, where the proximity effect creates dissipationless transport between superconductors connected by a normal material~\cite{Tinkham}. Our experimental data suggest the formation of superconductivity in $\mathrm{WTe_2}$ above Pd leads, as sketched in Fig.~\ref{WTe02}(c). These superconducting regions induce proximity effect in $\mathrm{WTe_2}$ between the leads, leading to the Josephson effect in the shorter junctions.

The observation of superconductivity may not be surprising, since $\mathrm{WTe_2}$ is known to become superconducting at different conditions: under pressure~\cite{presSC1,presSC2}, electron doping~\cite{dopSC} or electrostatic gating~\cite{gateSC1,gateSC2}. So superconductivity can occur in $\mathrm{WTe_2}$ on top of Pd due  to charge transfer~\cite{Shao} or due to flat-band formation in $\mathrm{WTe_2}$, as has recently been reported in another Weyl semimetal $\mathrm{Cd_3As_2}$~\cite{Shvetsov}. Another possibility is interdifussion of Pd and Te with the formation of superconducting $\mathrm{PdTe_x}$~\cite{PdTe,PdTe2} at the interface. To understand the reasons for superconductivity is beyond the scope of the current article, only the formation of Josephson junction within our samples is important.

\begin{figure*}[htb]
    \centering
    \includegraphics[width=1\textwidth]{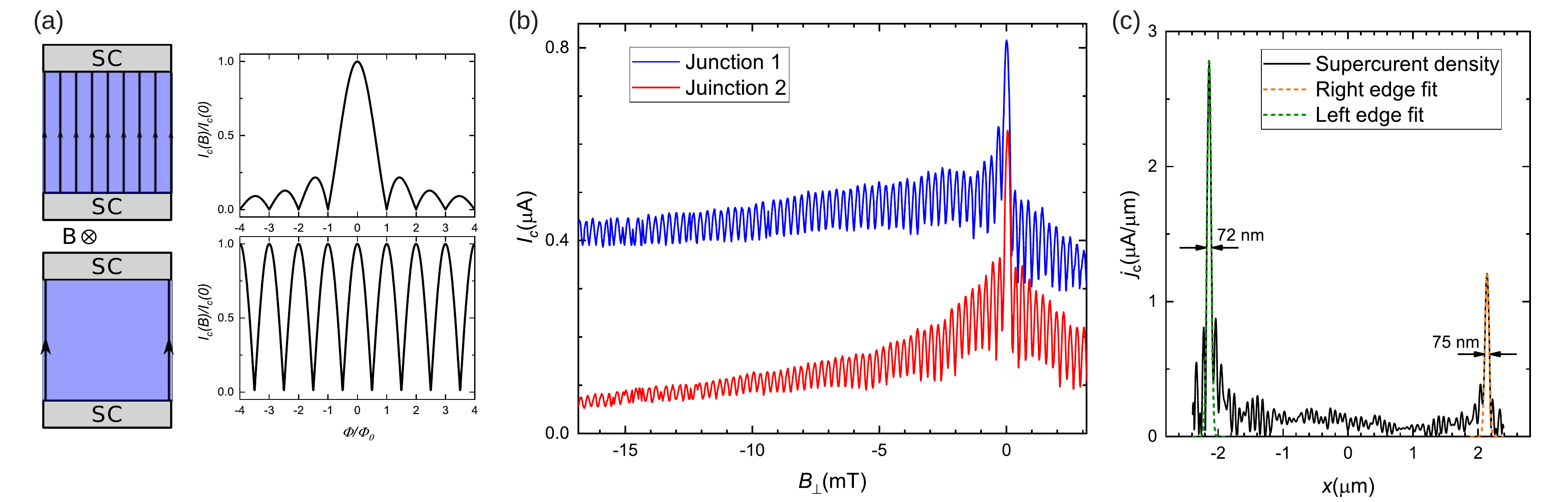}
    \caption{(\textbf{a}) Expected dependencies of the critical current of a 2D Josephson junction on $B_\perp$ for two different supercurrent distributions: for a uniform current distribution, $I_c(B_{\perp})$ shows rapidly decaying oscillations (Fraunhofer behaviour), whereas for two narrow edge states, the $I_c(B_{\perp})$ oscillations do not (or only weakly) decay in amplitude (SQUID behavior). (\textbf{b}) Critical current $I_c(B_{\perp})$ of junctions $1$ and $2$ as a function of $B_{\perp}$. A combination of a SQUID- and Fraunhofer-like behavior is observed, indicating a significant amount of edge supercurrent. (\textbf{c}) Supercurrent density distribution of junction~2 extracted from $I_c(B_\perp)$. Two distinctive edge states, each having a width of $\sim75~\si{\nano\meter}$, are observed.}
    \label{WTe03}
\end{figure*}

We can use the observed Josephson effect to obtain information about the current distribution in the $\mathrm{WTe_2}$ devices. The spatial current distribution defines the evolution of the critical current as a function of the flux through the Josephson Junction (JJ). When the supercurrent is uniformly distributed through the JJ, the critical current $I_c(B_{\perp})$ as a function of perpendicular magnetic field $B_{\perp}$ shows oscillations with a rapidly decaying amplitude (top in Fig.~\ref{WTe03}(a)). The central lobe is twice wider than the other lobes. This dependence of $I_c(B_{\perp})$ is known as the Fraunhofer pattern. If, on the other hand, the supercurrent flows only along the sample edges, as indicated in Fig.~\ref{WTe03}(a), $I_c(B_{\perp})$ displays slowly decaying oscillations typical for SQUIDs. The period of oscillations corresponds to a single flux quantum $\Phi_0=h/2e$  through the area enclosed by the SQUID~\cite{Tinkham}.

Fig.~\ref{WTe03}(b) shows the measured $I_c(B_{\perp})$ dependence for the two $1~\si{\micro\meter}$ long JJs. The critical current oscillates with perpendicular magnetic field. The central peak of $I_c$ has a width between one and two oscillations periods. The amplitude of the oscillations is decaying faster at smaller fields and slower at larger ones. The measured $I_c(B_{\perp})$ is a combination of a Fraunhofer pattern creating a peak of critical current at zero magnetic field and a SQUID-like pattern with more than $50$ visible oscillations. The period of these oscillations $\Delta B~\sim0.27~\si{\milli\tesla}$ is given by a flux $\Phi_0$ through the effective area of the junction $S_{eff}$. From $S_{eff}$ we obtain an effective junction length $L_{eff}=S_{eff}/W=1.75~\si{\micro\meter}$, where $W\sim4.3~\si{\micro\meter}$ is the sample width. $L_{eff}$ is larger than the length of the junction $L$ due to the penetration of magnetic field into the superconducting leads. A coexistence of the SQUID and Fraunhover behavior indicates the precence of edge and bulk supercurrent. The latter can be carried by the bulk of the crystal or by Fermi arc surface states~\cite{Kononov}. A persistence of the SQUID-like oscillations in magnetic field means that the edge supercurrent is carried by very narrow states.

To obtain the spatial distribution of the supercurrent, we performed a Fourier transform of $I_c(B_{\perp})$ by following the Dynes-Fulton approach~\cite{Dynes}. This method assumes a sinusoidal CPR and a nearly symmetric supercurrent distribution across the width of the junction. In this case the minima of $I_c(B)$ should approach zero. The result of the Fourier transform should therefore be more accurate for junction~2 as compared to junction~1, since the $I_c(B)$ minima are found to be much closer to zero in junction~2. Fig.~\ref{WTe03}(c) shows the result of the transformation for junction~2. The supercurrent peaks are very narrow, suggesting a strong edge confinement. The width of these supercurrent density peaks obtained from the Gaussian fit is below $80~\si{\nano\meter}$.

\begin{figure}[htb]
    \centering
    \includegraphics[width=0.4\textwidth]{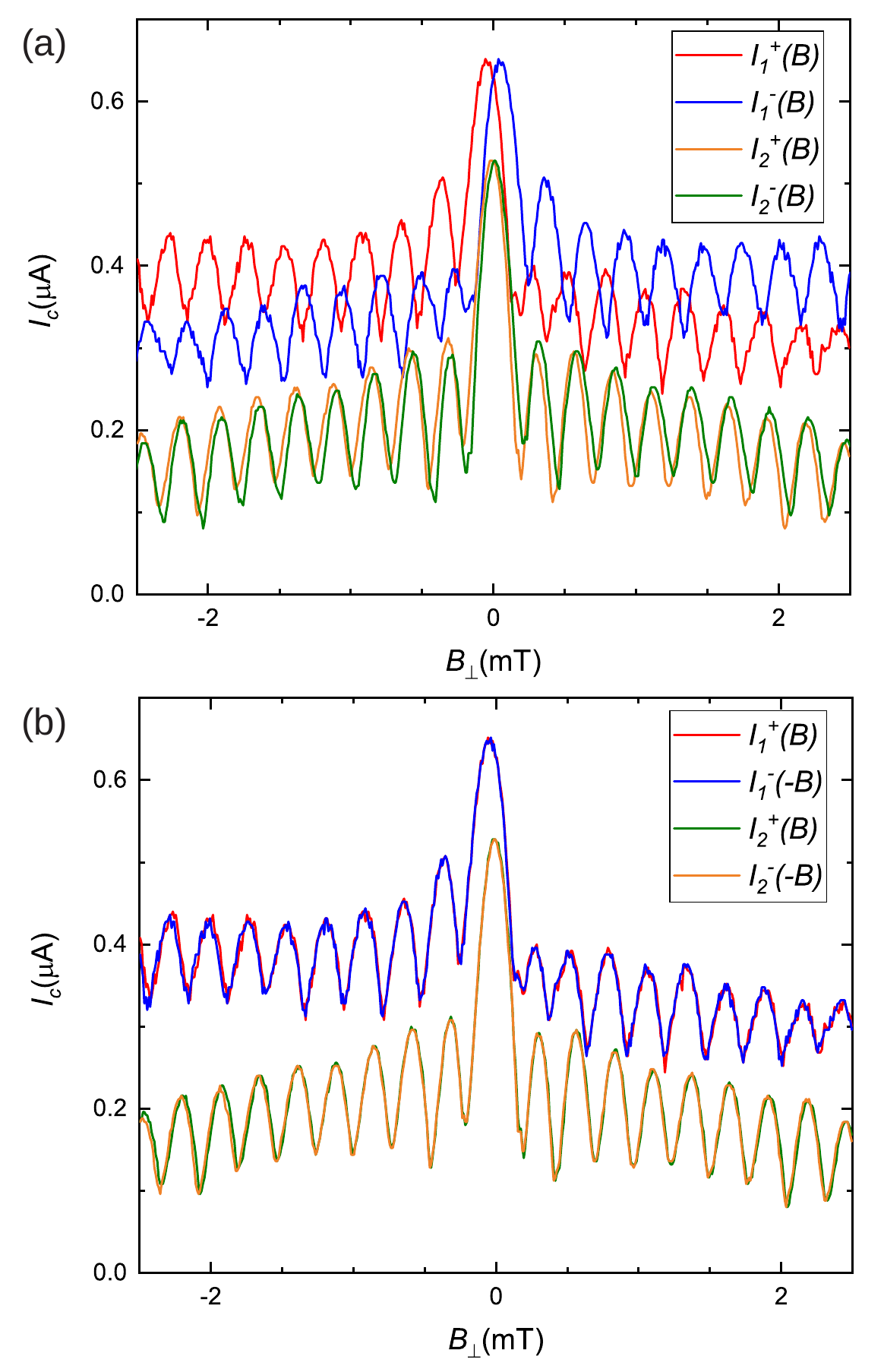}
    \caption{(\textbf{a}) Critical currents of two $1~\si{\micro\meter}$ long junctions of device~1 as a function of perpendicular magnetic field for positive $I_{1,2}^+$ and negative $I_{1,2}^-$ currents. $I_c(B_{\perp})$ lacks the symmetry to the change of current direction. (\textbf{b}) Same data as in (a) but with a reversed magnetic field for negative currents. Symmetry is preserved when both current and magnetic field are reversed.}
    \label{InvBreak}
\end{figure}

There is another reason, beyond an asymmetric current distribution~\cite{Dynes}, why the oscillations in Fig.~\ref{WTe03}(b) are not reaching zero: this can be caused by a non-sinusoidal CPR~\cite{Kurter,Pribiag} of the edge states. We can immediately confirm that the CPR is non-sinusoidal. This is seen as follows: the ratio of the critical currents of the two edge states $I_c^H/I_c^L$ is obtained from the ratio of the average critical current to the critical current oscillation amplitude $I_c^H/I_c^L=(I_c^{max}+I_c^{min})/(I_c^{max}-I_c^{min})$. For junction~1, this ratio is large, $\sim7\gg1$, hence, corresponding to a highly asymmetric SQUID. In such an asymmetric SQUID, the dependence of $I_c(B_{\perp})$ on $B_{\perp}$ mimics directly the CPR of the edge state with the lower critical current~\cite{Nanda}. But $I_c(B)$ for junction~1 is clearly not a sine function, as one can see from Fig.~\ref{WTe03}(b)and Fig.~\ref{InvBreak}. 

Additional evidence for a non-sinusoidal CPR can be obtained by looking at the symmetry of the dependence $I_c(B_{\perp})$ as a function of $B_{\perp}$.  For a conventional Josephson junction with a sinusoidal CPR, $I_c(B_{\perp})$ should be symmetrical with respect to current reversal $I^+_c(B_{\perp})=I^-_c(B_{\perp})$ and magnetic field reversal $I_c(-B_{\perp})=I_c(B_{\perp})$. Two requirements to break these symmetries are a non-sinusoidal CPR and an asymmetry in the current distribution~\cite{AJE}. However, the time-reversal symmetry conserves $I_c$ upon simultaneous reversal of the magnetic field and the current $I^{\pm}_c(B_{\perp})=I^{\mp}_c(-B_{\perp})$.

As is apparent from Fig.~\ref{InvBreak}(a), $I^{\pm}_c(B_{\perp})$ breaks  the symmetries both with current and field reversal. The symmetry is restored when the current and magnetic field are reversed simultaneously, as illustrated in Fig.~\ref{InvBreak}(b). The time-reversal symmetry allows us to exclude flux trapping in the JJ~\cite{Golod} as a reason for the observed asymmetries. The asymmetries in Fig.~\ref{InvBreak} match the prediction of AJE~\cite{AJE} and require a non-sinusoidal CPR and an asymmetry in current distribution.

\begin{figure*}[htb]
    \centering
    \includegraphics[clip, trim=0cm 10cm 0cm 10cm,width=1\textwidth]{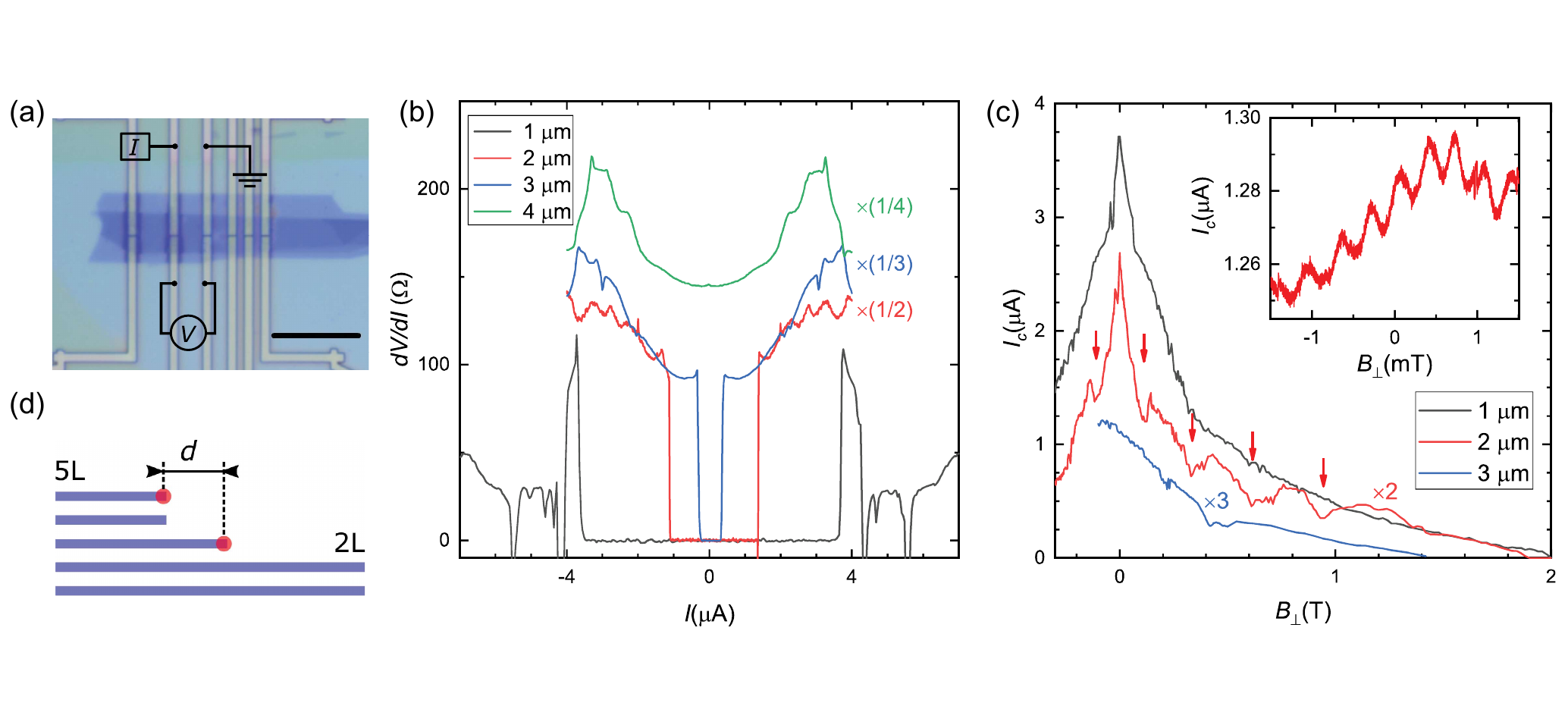}
    \caption{(\textbf{a}) Optical image of device~2 (scale bar $10~\si{\micro\meter}$) with a sketch of the measurement setup for a single junction. Each Pd lead has a $100~\si{\nano\meter}$ gap in the middle, which is located below the thicker part of the $\mathrm{WTe_2}$ flake. The gaps split every Pd lead into two independent normal contacts to the common superconducting region. (\textbf{b}) Four-terminal $dV/dI$ of junctions with different lengths divided by the length of the corresponding junctions in $\si{\micro\meter}$. The Josephson effect is present in junctions that are up to $3~\si{\micro\meter}$ long. (\textbf{c}) Critical current $I_c(B_{\perp})$ as a function of perpendicular magnetic field $B_{\perp}$. (Note: $I_c$ is here multiplied by the length of the corresponding junctions in $\si{\micro\meter}$). The arrows highlight the periodic low-frequency modulation of $I_c$  for the $2~\si{\micro\meter}$ long junction. Inset: $I_c(B_{\perp})$ for the $2~\si{\micro\meter}$ long junction zoomed in to the small magnetic field region. A fast periodic oscillation with an amplitude of $\sim 1\,\%$ is clearly discerned. (\textbf{d}) Sketch of a cross section of the sample near the step from 5L to 2L, illustrating the possibility that multiple 1D channels along the step appear.}
    \label{WTe01}
\end{figure*}

We have found before that the supercurrent in few-layer $\mathrm{WTe_2}$ is of 1D nature, flowing predominately along the edges and has a non-sinusoidal CPR. With the next sample we demonstrate that 1D conducting states can also reside at step edges of $\mathrm{WTe_2}$ and they are remarkably robust. Device~2, shown in Fig.~\ref{WTe01}(a), is as before a hBN-covered few-layer $\mathrm{WTe_2}$ flake placed on top of Pd leads. The main difference from the previous device is non-uniform thickness: the middle part is 5 layer thick and the outer parts are bilayers. The low temperature conductivity in $\mathrm{WTe_2}$ diminishes with decrease in number of layers with the bilayer being an insulator~\cite{Fei, Wang}.

Fig.~\ref{WTe01}(b) shows $dV/dI(I)$ traces for different junctions normalized by the length of the junction. The differential resistance goes to zero for $1-3~\si{\micro\meter}$ long junctions, indicating the presence of Josephson current. The normal state resistance per unit length is comparable for all junctions, yielding $\sim100~\si{\ohm\per\micro\meter}$. For this sample, the product $I_c R_N\sim150-380~\si{\micro\volt}$ depending on the junction and the way the normal state resistance $R_N$ is defined. This value is close to the theoretical prediction for a short ballistic Josephson junction: $I_c R_N=\pi\Delta/e\sim540~\si{\micro\volt}$~\cite{Kulik}. Here, we estimate the energy gap following the formula $\Delta(T=0)=1.76k_B T_c$~\cite{Tinkham} with $T_c=1.1~\si{\kelvin}$ defined as the maximal temperature where signs of superconductivity in the samples are still present. The agreement between the $I_c R_N$ product and the theoretical value implies that there is a strong proximity effect and the JJs are close to the short ballistic limit.

The Josephson current for all junctions survives magnetic fields above $1~\si{\tesla}$, see Fig.~\ref{WTe01}(c). This is inconsistent with a uniform supercurrent, since even for the shortest junction it would correspond to $B S/\Phi_0\sim 2000$ flux quantum through the JJ area. A robust large field supercurrent implies highly localized 1D channels that carry the supercurrent. The only possible place for these states are the steps from the five-layer part to bilayers, since the bilayer itself does not conduct. At a closer look, oscillations of $I_c(B_{\perp})$ are visible for the $2~\si{\micro\meter}$ long junction, see the inset to Fig.~\ref{WTe01}(c). The oscillations are clearly of a SQUID character with a period $\Delta B~\sim0.33~\si{\milli\tesla}$. This period yields a smaller area $S=\Phi_0/\Delta B\sim6.1~\si{\micro\meter\squared}$ than the relevant junction’s area $9~\si{\micro\meter\squared}$. This mismatch is likely a consequence of the sample geometry and discussed in more detail in the supplementary.

The measurement of $I_c(B_{\perp})$ of the $2~\si{\micro\meter}$ long junction shows additional oscillations with a larger period of $\delta B\sim~0.3~\si{\tesla}$ (red arrows in Fig.~\ref{WTe01}(c)).  Similar oscillations were previously observed for topological hinge states in bismuth and were linked to a difference in wavevectors of electrons and holes forming the Andreev pairs~\cite{Li}. The observed period of oscillations is in agreement with the expected value $\delta B\sim2\pi\hbar v_{F}/g_{eff}\mu_{B}L\sim0.15-0.7~\si{\tesla}$, where $L=2~\si{\micro\meter}$ is the length of the junction, $v_F\sim2\cdot10^5~\si{\meter\per\second}$~\cite{Huang1} the Fermi velocity and $g_{eff}\sim10-50$~\cite{Bi} the Land\'e  \emph{g}-factor. Alternatively, a slower oscillation could reflect the presence of multiple states on terraces from 5 layers to the bilayer, as illustrated in Fig.~\ref{WTe01}(d). The width $d$ of this region can be estimated from the ratio of the periods of the slow $\delta B\sim0.3~\si{\tesla}$ and fast oscillations $\Delta B\sim0.33~\si{\milli\tesla}$ and the width of the junction $W\sim{4.5~\si{\micro\meter}}$: $d\sim W\delta B/\Delta B\sim5~\si{\nano\meter}$. This value is an upper estimate for the width of the edge states.

The observation of strong Josephson coupling through 1D edge states with non-sinusoidal CPR suggests a topological origin of these states~\cite{Sochnikov}. The only predicted 1D topological states in few-layer $\mathrm{WTe_2}$ are hinge states of a HOTI~\cite{HOTI}. We think that this is very plausible, since our data reproduces many features previously observed in bismuth, which is a HOTI~\cite{Murani}. However, there are still some open questions. Currently we can not resolve, if the states are indeed residing on opposite hinges as expected in a HOTI. Also the critical current values are higher, than expected for a single ballistic channel  $I_c^{1D}=\pi\Delta/eR_k=e\Delta/2\hbar\sim20~\si{\nano\ampere}$. This discrepancy is also present in bismuth, and can be accounted by multiple states at several terraces on the edges and degeneracy of edge states due to multiple orbitals~\cite{Murani}.

In conclusion, we present an experimental study of Josephson transport in encapsulated few-layer $\mathrm{WTe_2}$ samples. Our data strongly suggest the  presence of 1D states residing on steps and edges of $\mathrm{WTe_2}$. The Josephson currents in these 1D states are extremely robust. They survive magnetic fields up to $2~\si{\tesla}$ and extend over distances up to $3~\si{\micro\meter}$. Moreover, the supercurrent demonstrates signs of non-sinusoidal CPR. Our findings fit well with the recent prediction of higher-order topological insulator states in $\mathrm{WTe_2}$~\cite{HOTI} and demonstrate many features previosly observed only in another HOTI - bismuth~\cite{Murani, Li}.

\textit{Note:} During the preparation of this manuscript we became aware of two recent preprints~\cite{Choi, Huang} demonstrating edge transport in $\mathrm{WTe_2}$ obtained by the proximity effect from superconducting Nb leads. The experimental results in these preprints are in good agreement with our conclusions. In comparison to the former, our samples are in the thin limit and they additionally demonstrate a stronger Josephson coupling over longer distances. They thereby provide a more compelling evidence for Josephson coupling through highly localized narrow 1D states residing on the steps of $\mathrm{WTe_2}$.

\section*{Acknowledgments}
We thank D.~Indolese for the help with measurements of the critical current and fruitful discussions, M.~Endres for his help with the exfoliation and identification of $\mathrm{WTe_2}$ flakes, M.~Joodaki for her contribution to the optical identification of $\mathrm{WTe_2}$ flakes thickness and A.~Baumgartner for fruitful discussions. A.K. was supported by the Georg H.~Endress foundation. This project has received further funding from the European Research Council (ERC) under the European Union’s Horizon 2020 research and innovation programme: grant agreement No 787414 TopSupra, by the Swiss National Science Foundation through the National Centre of Competence in Research Quantum Science and Technology (QSIT), and by the Swiss Nanoscience Institute (SNI).
K.W. and T.T. acknowledge support from the Elemental Strategy Initiative conducted by MEXT, Japan and the CREST (JPMJCR15F3), JST.
D.M. and J.Y. acknowledge support from the U.S. Department of Energy (U.S.-DOE), Office of Science - Basic Energy Sciences (BES), Materials Sciences and Engineering Division.

\section*{Authors contributions}
A.K. fabricated the devices 1, 2, performed the measurements and analyzed the data. G.A. optimized the fabrication recipe, developed the thickness determination method by optical contrast and together with A.K. fabricated and measured device S1. K.Q., J.Y. and D.M. provided $\mathrm{WTe_2}$ crystals. K.W. and T.T. provided hBN crystals. A.K prepared the manuscript. C.S. initiated and supervised the project and participated in all discussions. All authors contributed to the manuscript.

\section*{Data availability}
All data in this publication are available in numerical form in the Zenodo repository at \url{https://doi.org/10.5281/zenodo.3526560}.

\bibliographystyle{abbrv}


\clearpage
\widetext
\setcounter{equation}{0}
\setcounter{figure}{0}
\setcounter{table}{0}
\setcounter{page}{1}
\setcounter{section}{0}

\makeatletter
\renewcommand{\thefigure}{S\@arabic\c@figure}
\makeatother

\renewcommand{\title}{Supplementary information of\\
One-dimensional edge transport in few-layer $\mathrm{WTe_2}$}
\begin{center}
\textbf{\large Supplementary information of\\
One-dimensional edge transport in few-layer $\mathrm{WTe_2}$}
\end{center}
\renewcommand{\author}{}
\renewcommand{\email}{}
\renewcommand{\affiliation}{}
\renewcommand{\bibnumfmt}[1]{[S#1]}
\renewcommand{\citenumfont}[1]{S#1}

\section*{S1. Methods}	

\subsection*{Fabrication}
Contacts were patterned by a standard e-beam lithography on $p$-doped Si substrates with $295~\si{\nano\meter}$ thick $\mathrm{SiO_2}$ layer on top. $3~\si{\nano\meter}$ of titanium and $12~\si{\nano\meter}$ of palladium were deposited in an e-beam evaporator system, followed by lift-off in hot acetone. hBN flakes were mechanically exfoliated on similar substrates under ambient conditions and $10-30~\si{\nano\meter}$ thick flakes without visible steps and signs of residues were preselected. $\mathrm{WTe_2}$ flakes were exfoliated from flux grown $\mathrm{WTe_2}$~\cite{growth} in a $\mathrm{N_2}$ filled glovebox with an oxygen level below $0.5~\mathrm{ppm}$. We optically identified thin (below 15 layers) elongated flakes that are oriented along the $a$-axis without visible steps, except for device~3. The thickness of $\mathrm{WTe_2}$ flakes was identified using the optical contrast method~\cite{Blake}. We first picked up an exfoliated hBN flake using the polymer dry transfer technique~\cite{transfer}. This stamp was then used to pick up selected $\mathrm{WTe_2}$ flakes and place the stack on the prepatterned leads. Thus, $\mathrm{WTe_2}$ flakes were always protected from oxidation, initially by  keeping the exfoliated flakes in the oxygen-free environment of the glovebox and later by the hBN cover.

\subsection*{Measurements}
The low temperature measurements were done in a dilution refrigerator with a base temperature of $30~\si{\milli\kelvin}$. The insert of the cryostat was fitted with low temperature line filters. Additional $10~\si{\nano\farad}$ $\pi$-filters were attached at room temperature. We determined the differential electrical resistance by current biasing the sample with both DC and AC components and measuring the voltage over the sample using an $SR-830$ lock-in amplifier. The DC current is obtained from a voltage source connected to the device through a series resistor with a resistance value of $100~\si{\kilo\ohm}$. The AC component is added through a transformer. The voltage over the sample was amplified by an in-house built low noise differential amplifier. We used AC frequencies ranging from $77~\si{\hertz}$ to $277~\si{\hertz}$. All measurements were done in the linear response regime with an AC excitation current below $4~\si{\nano\ampere}$.

For device~2 with a high $I_c R_n$ product of $\sim200~\si{\micro\volt}$ we also employed statistical measurements of the switching current to obtain $I_c$. We bias the sample with a time-dependent current ramp for which the current increases at a constant rate. The time before the JJ switching from the superconducting to the normal state is measured with a counter, which is stopped by a trigger signal obtained from the sharp increase of the voltage across the junction from $0$ to $\sim I_c R_n$. This time is averaged over 200 current ramps and used as the value for the critical current $I_c$. In reality, the switching current is smaller than the "true" critical current $I_c$. To set the current, we used a signal generator creating a saw-tooth signal at frequencies between $177$ and $277~\si{\hertz}$ connected through a $10~\si{\kilo\ohm}$ resistor in series with the sample. The voltage drop across the junction was amplified $1000$ times before reaching the trigger input of the counter set to a threshold value of $15~\si{\milli\volt}$.

\section*{S2. Superconductivity induced in $\mathrm{WTe_2}$ by normal leads}

\begin{figure}[h!]
    \centering
    \includegraphics[clip, trim=0cm 5.5cm 0cm 5.5cm, width=1\textwidth]{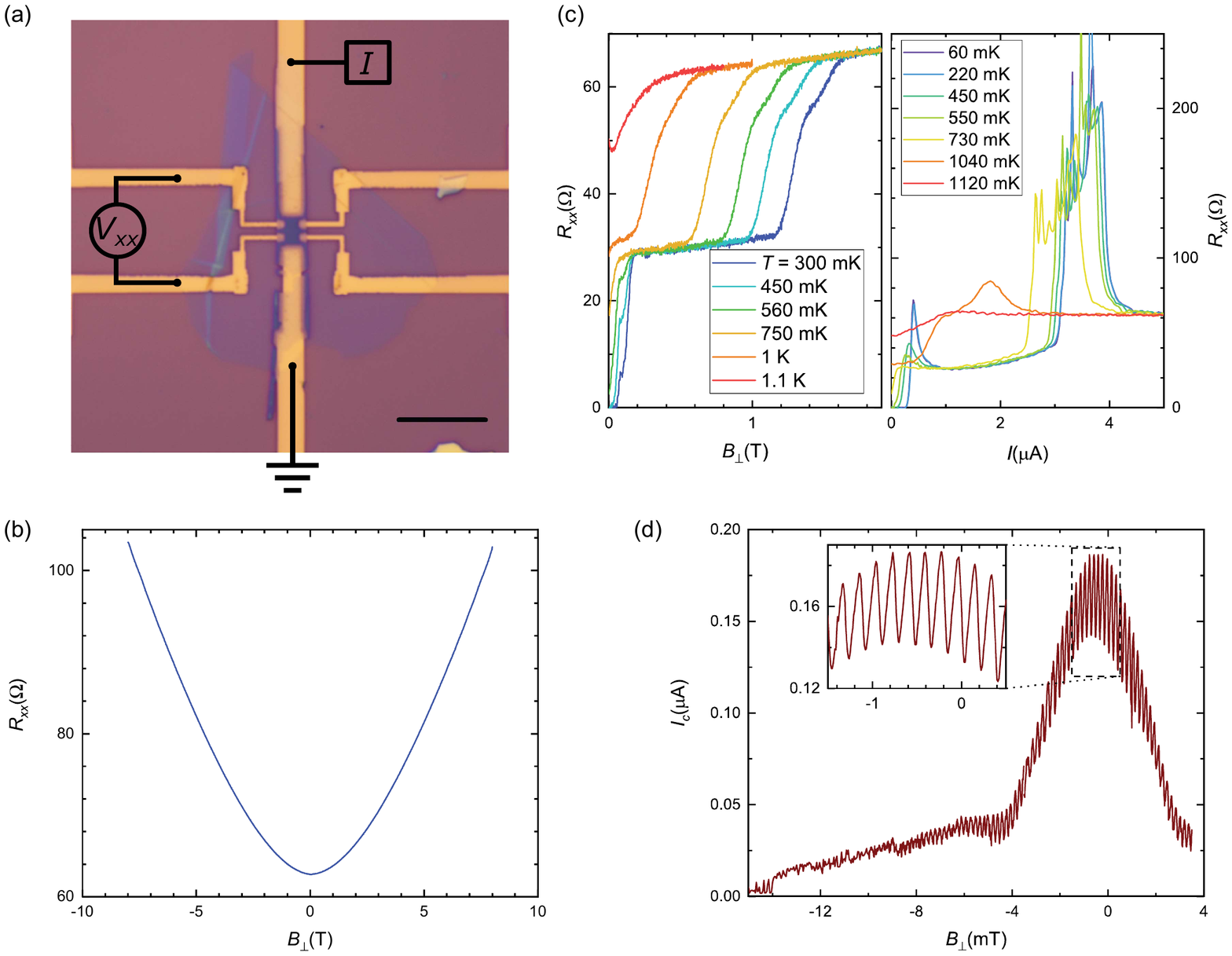}
    \caption{\textbf{Superconductivity and Josephson effect in a few-layer $\mathrm{WTe_2}$ device in a Hall bar geometry.} (\textbf{a}) Optical image of the device~1 (scale bar $10~\mu m$) with a sketch of the four-terminal measurement setup. (\textbf{b}) Longitudinal resistance of device~1 as a function of perpendicular magnetic field (along the c-axis) measured at $4~\si{\kelvin}$. The magnetoresistance does not saturate at high magnetic fields. (\textbf{c}) The longitudinal differential resistance $R_{xx}$ of the device as a function of perpendicular magnetic field $B_{\perp}$ (left) and current $I$ (right) at different temperatures. $R_{xx}$ assumes a zero value below a certain temperature, magnetic field or current value. This is a clear sign of superconductivity in the sample. (\textbf{d})  The critical current $I_c(B_{\perp})$ as a function of perpendicular magnetic field $B_{\perp}$ measured at $60~\si{\milli\kelvin}$. $I_c$ is extracted from the position where the two-terminal differential resistance $R(I)$ as a function of bias current shows a sharp increase, associated with the transition to the intermediate state.  The inset shows the oscillation of the critical current in more details. The period of oscillations is $0.2~\si{\milli\tesla}$. }
    \label{WTeS1}
\end{figure}

A few-layer $\mathrm{WTe_2}$ device covered with hBN and shaped in a Hall bar geometry has been fabricated by placing a stripe-shaped $\mathrm{WTe_2}$ flake on prepatterned Pd leads, as demonstrated in Fig.~\ref{WTeS1}(a). We measured the longitudinal resistance $R_{xx}$ of the $\mathrm{WTe_2}$ flake in a four-probe setup. At $4~\si{\kelvin}$ the device demonstrates a non-saturating magnetoresistance in perpendicular magnetic field $B_{\perp}$ up to $8~\si{\tesla}$ (Fig.~\ref{WTeS1}(b)), which is a signature of a high quality $\mathrm{WTe_2}$ crystal~\cite{WTe_qual}. The relatively small magnetoresistance can be explained by the small ($\sim7$ layers) $\mathrm{WTe_2}$ thickness~\cite{Xiang}.

Upon cooling down to below $1.1~\si{\kelvin}$, the behavior of $R_{xx}$ changes drastically (Fig.~\ref{WTeS1}(c)): at low magnetic fields the resistance is zero, then there is a rapid transition where $R_{xx}$ increases to an intermediate state with approximately half of the $4~\si{\kelvin}$ resistance value. At even higher magnetic fields, a second rapid increase occurs where the normal state resistance of the device is restored. Besides, both increases of the resistance move towards zero field when the temperature is increased. The zero $R_{xx}$ state disappears completely above $600~\si{\milli\kelvin}$ and the intermediate resistance disappears above $1.1~\si{\kelvin}$. A similar behavior is obtained at zero magnetic field as a function of current: $R_{xx}$ first switches to the intermediate state, followed by a switch to the normal-state resistance. This behavior is typical for Josephson junctions, where the change of the resistance from zero to a finite value corresponds to  the disappearance of the  Josephson current and the second increase of the resistance reflects the transition to the normal state of the superconducting leads.

Our $\mathrm{WTe_2}$ crystals are not intrinsically superconducting. In all the samples we observe superconductivity only with shorter junctions. For a superconductor no dependence on the contacts separation is expected. We can also exclude the possibility that the Pd leads are superconducting and induce the superconductivity in $\mathrm{WTe_2}$ by the proximity effect. In this case, the interface resistance between Pd and $\mathrm{WTe_2}$ is expected to be zero, but in our samples this resistance is measured to be $\sim500~\si{\ohm}$.

Fig.~\ref{WTeS1}(d) displays the critical current of the device shown in Fig.~\ref{WTeS1}(a). The critical current is determined by the current value where the two-terminal resistance $R$ jumps from the zero-state to the intermediate-state value discussed before and shown in Fig.~\ref{WTeS1}(b). The $I_c(B_{\perp})$ dependence is a convolution of a SQUID-like behavior with many rapid oscillations and a Fraunhofer pattern with a much lower frequency. The period of the fast oscillation $\triangle B\sim0.2~\si{\milli\tesla}$ roughly corresponds to a single flux quantum $\Phi_0$ through the area $S$ of the sample ($S~\sim12~\si{\micro\meter\squared}$), indicating that the supercurrent flows along the edges of the $\mathrm{WTe_2}$ flake. The Fraunhofer shape of the envelope of oscillations reflects the finite width of states hosting the supercurrent.

\section*{S3. Supercurrent distribution in the device~2.}
\begin{figure}[htb]
    \centering
    \includegraphics[width=0.8\textwidth]{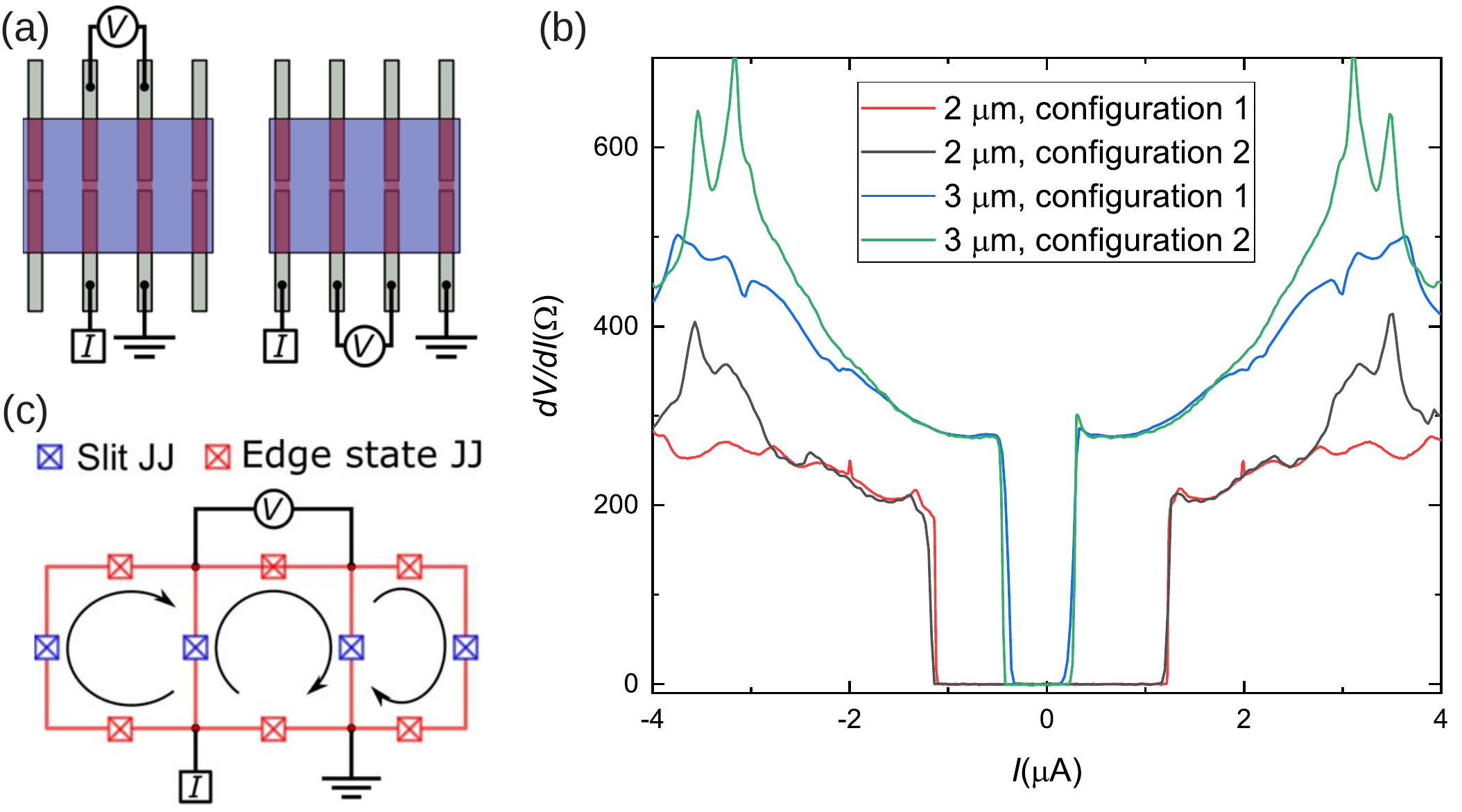}
    \caption{(\textbf{a}) Schematics of the two measurement setups. Left: the current is passed through two neighboring leads, while the voltage is measured across the two leads that reside on the opposite side and are separated by a small gap. Right: the current is injected through two leads outside the studied junction, one on the left and the other on the right side, while the voltage is measured across the studied junction. Similar results are obtained with both setups. (\textbf{b}) Comparison of $dV/dI(I)$ dependencies obtained with different measurements setups. (\textbf{c}) Electrical circuit that resembles the main current distribution in device 2. Only junctions neighboring to the studied junction are shown. There are two different kinds of JJs: edge states between SC regions above the Pd leads form JJs (red), but also the short slits in the Pd leads define JJs (blue). When a current is passed between a pair of Pd leads, the supercurrent is redistributed across the network of JJs.}
    \label{Steps}
\end{figure}

In device~2 each Pd lead is split into two by a $100~\si{\nano\meter}$ gap. The two Pd leads formed by the slit are acting as two normal contacts to a common superconducting region. This provides an opportunity to measure the resistance in a four-probe manner using only contacts on the studied junction, as shown by the left schematic in Fig.~\ref{Steps}(a). This is different from the measurements employed in the device~1, as depicted in the right schematic. We made sure that such measurements are correct through the direct comparison of $dV/dI(I)$ obtained for JJs, where both types of measurements are available. Examples of such measurements are provided in Fig.~\ref{Steps}(b). The $dV/dI(I)$ dependencies are quite similar at smaller currents, but diverging at high currents when the superconductivity is suppressed.

The presence of gaps in the Pd leads can complicate the interpretation of $I_c(B_{\perp})$, since the $100~\si{\nano\meter}$ wide slits in the Pd leads may act as additional JJs. Together with the JJs formed by the edge states, a network of JJs is formed, as schematically shown in Fig.~\ref{Steps}(c). In this case, the distribution of supercurrent across the whole network defines $I_c(B_{\perp})$ for each pair of Pd leads. We think that this is the reason for the oscillation period mismatch in the $2~\si{\micro\meter}$ long junction. As discussed in the main text, the period yields a smaller area $S=\Phi_0/\Delta B\sim6.1~\si{\micro\meter\squared}$ than the relevant junction’s area $9~\si{\micro\meter\squared}$. However, it is comparable with the area of the neighboring $1~\si{\micro\meter}$ long JJ taking into account flux focusing: $L_{eff}=(L+2\lambda_{L})=S/W=1.4~\si{\micro\meter}$.

The formation of a network of JJs can complicate the observation and the interpretation of $I_c(B)$ oscillations, but does not affect the conclusion that 1D supercurrent carrying states are present at the steps of our flake. In this picture, a small amplitude of SQUID-like oscillations, as we observe here, would mean that the slit JJs are relatively weak compared to the JJs defined by the 1D transport channels. Although the slit junctions are shorter, their critical current must be small, suggesting that bulk states in $\mathrm{WTe_2}$ have a much smaller mobility than the 1D states.

\section*{S4. Differential resistance as a function of magnetic field and current for the device~2.}
\begin{figure}[h!]
    \centering
    \includegraphics[clip, trim=2cm 6.5cm 1.5cm 7cm, width=0.8\textwidth]{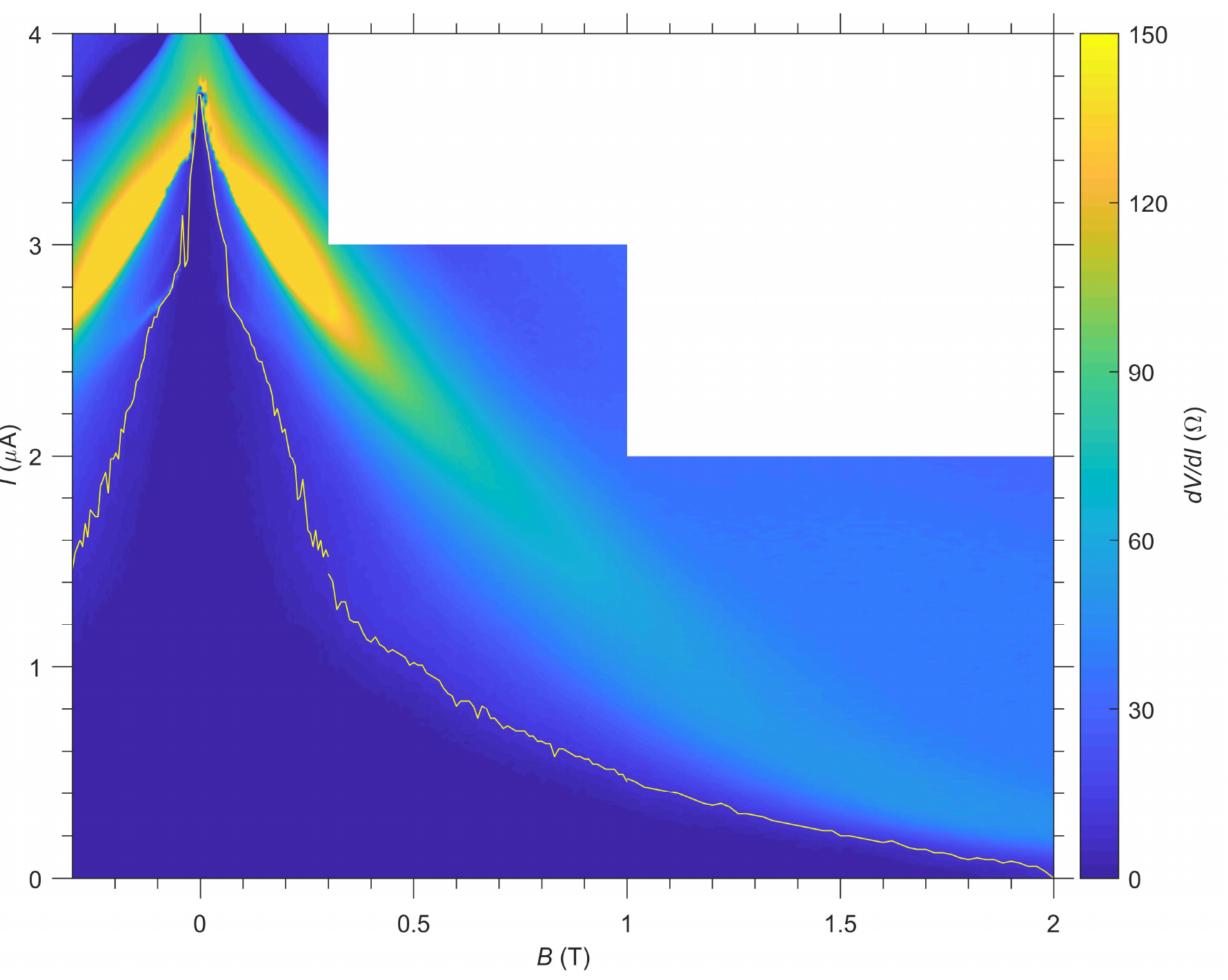}
    \caption{Differential resistance as a function of perpendicular magnetic field and current through the $1~\si{\micro\meter}$ junction. The yellow line indicates extracted $I_c(B)$.}
    \label{J1}
\end{figure}

\begin{figure}[h!]
    \centering
    \includegraphics[clip, trim=1.5cm 6.5cm 1.5cm 7cm, width=0.8\textwidth]{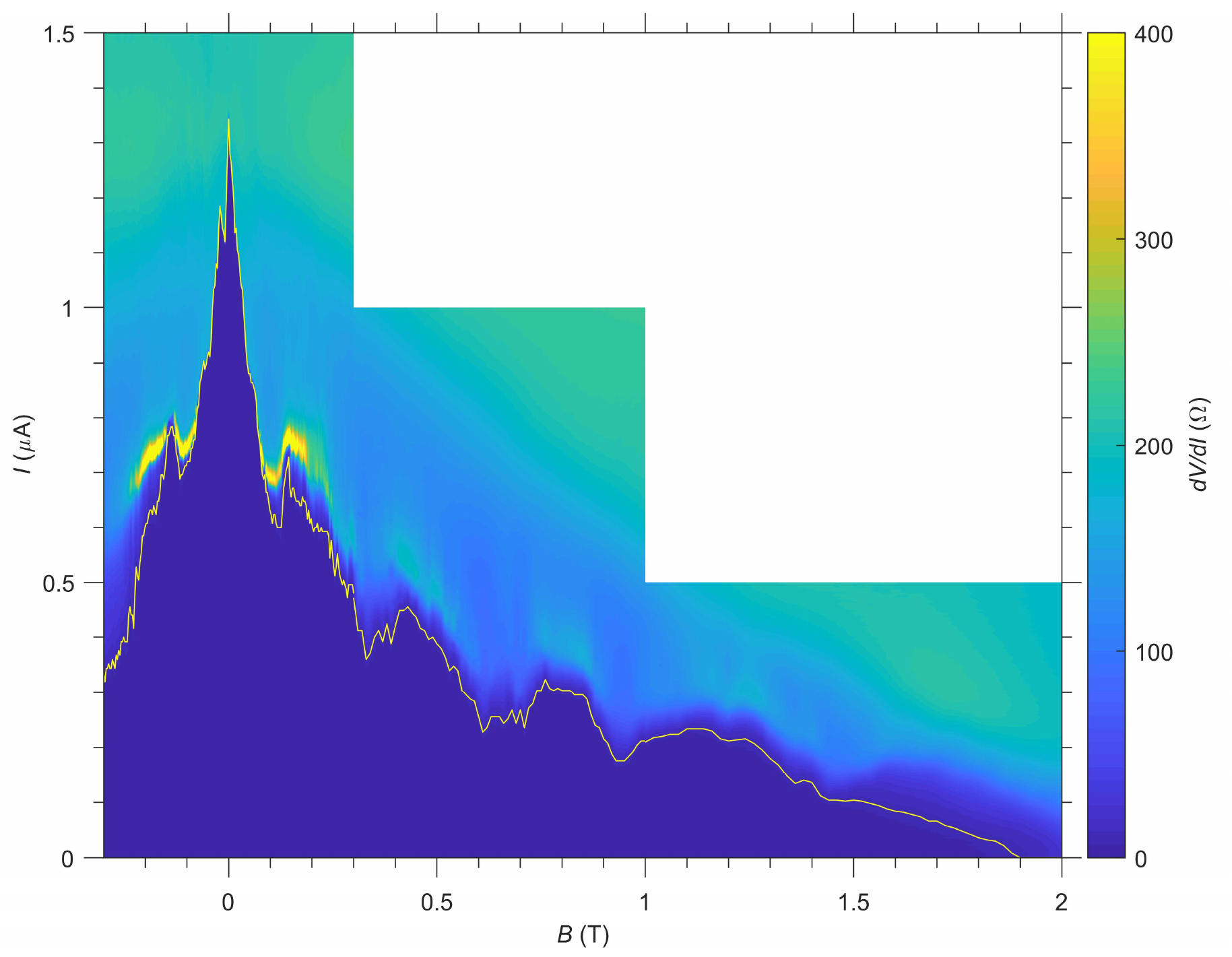}
    \caption{Differential resistance as a function of perpendicular magnetic field and current through the $2~\si{\micro\meter}$ junction. The yellow line indicates extracted $I_c(B)$.}
    \label{J2}
\end{figure}

\begin{figure}[h!]
    \centering
    \includegraphics[clip, trim=1.5cm 6.5cm 1.5cm 7cm, width=0.8\textwidth]{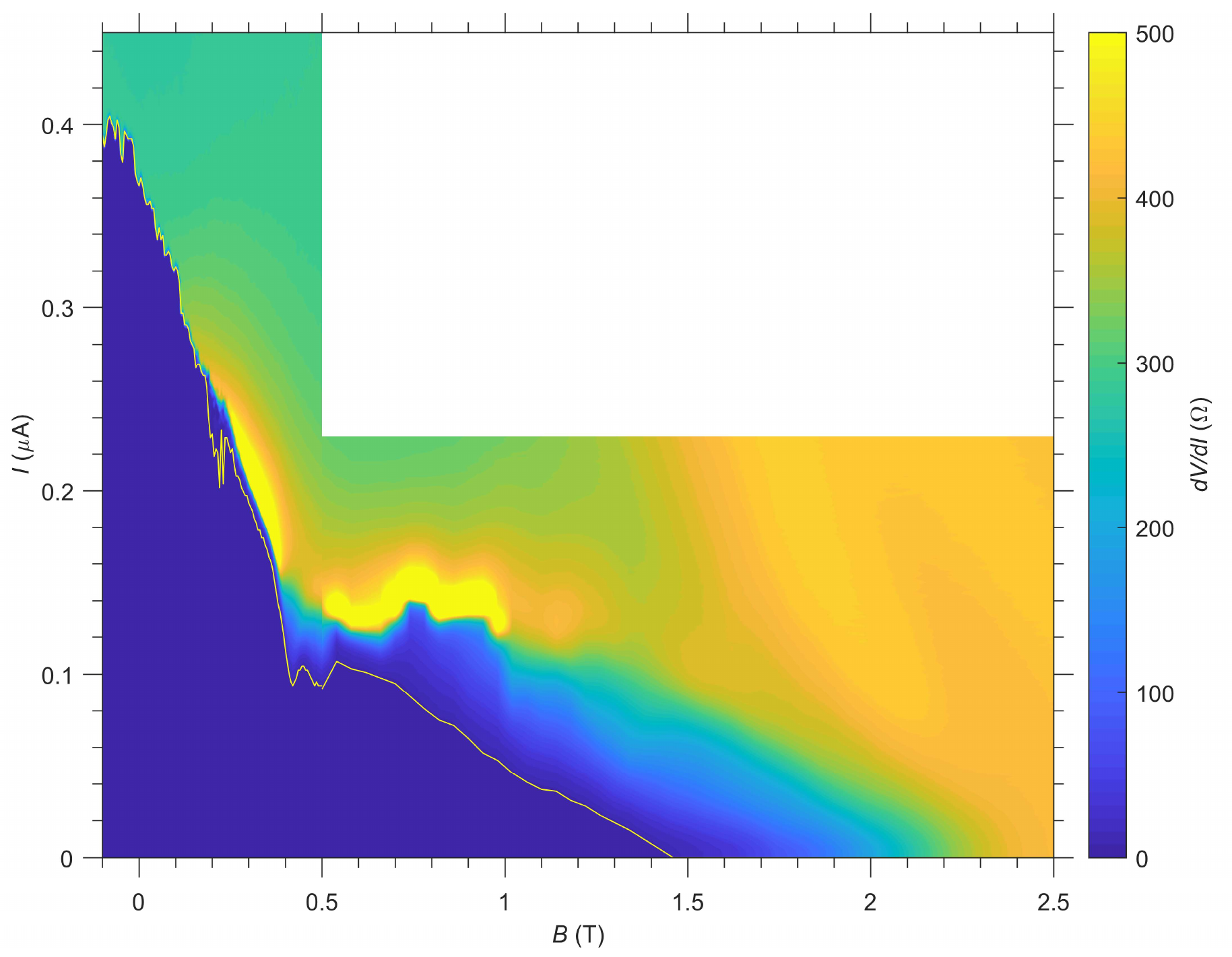}
    \caption{Differential resistance as a function of perpendicular magnetic field and current through the $3~\si{\micro\meter}$ junction. The yellow line indicates extracted $I_c(B)$.}
    \label{J3}
\end{figure}

\bibliographystyle{abbrv}

\end{document}